\documentstyle[12pt]{article}
\begin{document}

\begin{center}
{\bf {\large Regularization of the Density of States Fluctuation
Contribution in Magnetic Field}}
\end{center}

\vskip 1.3truecm

\begin{center}
{\bf A.I.Buzdin}

{Centre de Physique Theorique et de Modelisation,}

{Universit\`{e} Bordeaux I, CNRS-URA 1537,}

{33174 Gradignian, Cedex, France}

\vskip 0.5truecm

{\bf A.A.Varlamov}

{Forum, INFM, Dipartimento di Fisica, Universit\`{a} di Firenze,\\Largo
E.Fermi, 2 50125 Firenze, Italy}

{and}

{Department of Theoretical Physics, Moscow Institute for Steel and Alloys,\\%
Leninski pr. 4, Moscow 117936, Russia}
\end{center}

\begin{abstract}
The fit of the experimental data on c-axis magnetoresistance of HTS above
the transition temperature with the theory based on the fluctuation
renormalization of the one-electron density of states (DOS) is exellent in
weak magnetic fields but meets the noticible difficulties in the region of
strong fields. This is due to the formal divergency of the DOS contribution
to conductivity and the dependence of the cut-off parameter on the magnetic
field itself. We propose the scheme of the regularization of the problem.
This permits us to obtain the expression for the magnetic field dependent
part of DOS conductivity as a convergent serie independent on cut-off. We
also calculate analitically the asymptotics for all regions of magnetic
fields. The results demonstrate the robustness of the DOS contribution with
respect to\ the magnetic field effect: in strong fields ($H>H_{c2}(T-T_{c})$
) it decreases logarithmically only while Aslamazov-Larkin and anomalous
Maki-Thompson contributions diminish as powers of $\frac{H_{c2}}{H}$.
\end{abstract}


\newpage

The recently proposed idea of the importance of the one-electron density of
states (DOS) renormalization due to superconducting fluctuations was
successfully applied to the explanation of the c-axis resistance temperature
dependence, optical conductivity, c-axis magnetoresistance, NMR rate
behavior near $T_c$, tunneling conductance and other normal state anomalies
of HTS in metallic part of the HTS phase diagram \cite{VBML98}. It turns out
that the strong renormalization of the DOS in a very narrow vicinity of the
Fermi level ($E\sim T-T_c$) together with the particle conservation law
leads to the observation of the pseudogap-like phenomena in a set of normal
state characteristics of HTS. The prediction \cite{BMV95,BDKLV93} of the
sign change of the c-axis magnetoresistance was especially interesting and
later this effect was observed on several experiments \cite{Rapp96,Heine98}.
The fit of these experimental data with the theory based on the fluctuation
renormalization of the one-electron density of states \cite{ILVY93,BDKLV93}
is excellent for weak magnetic fields but meets the noticeable difficulties
in the region of strong fields. This is due to the formal divergency of the
DOS contribution to conductivity and the dependence of the cut-off parameter
on the magnetic field itself \cite{BDKLV93}. The similar problem arises for
the DOS contributions to NMR and other physical properties measured in
magnetic field.

In this communication we clarify the problem of the regularization of DOS
contribution in an arbitrary magnetic field on the example of c-axis
magnetoconductivity of quasi-two-dimensional superconductor looking forward
to apply the proposed theory to the measurements on HTS compounds. This
permits us to leave apart in this communication the discussion of the
Aslamazov-Larkin and anomalous Maki-Thompson contributions, which were
studied in details (see, for example, \cite{ahl89}) but due to the low
interlayer transparency of the most of HTS materials they may be omitted
beyond the immediate vicinity of $T_c$. Below we imply the model of electron
spectrum in the form of corrugated cylinder and use notations introduced in 
\cite{BDKLV93}, where the tensor of fluctuation conductivity of layered
superconductor in magnetic field was discussed. The magnetic field is
supposed to be oriented along c-axis, so finally the problem is reduced to
the carrying out of the summation over the Landau states of the center of
mass of fluctuation Cooper pair \cite{BDKLV93}.To avoid the problem of the
ultraviolet divergence of the DOS contribution with the badly defined
cut-off depending on the magnetic field \cite{BDKLV93} we calculate the
cut-off independent difference of magnetoconductivity \cite{BD96}in field
and its absence: 
\begin{eqnarray}
\Delta \sigma _c^{DOS}({h},\epsilon )=\sigma _c^{DOS}({h},\epsilon )-\sigma
_c^{DOS}(0,\epsilon ),  \label{delta}
\end{eqnarray}
where $\epsilon =\ln {T/T_c}=\frac{T-T_c}{T_c}$ is reduced temperature, $%
h=\frac H{H_{c2}(0)}$ is the dimensionless  magnetic field (both these
parameters are supposed to be small: $\epsilon ,h\ll 1$).

For this purpose the zero-field value $\sigma _c^{DOS}(0,\epsilon )$ \cite
{BDKLV93} may be rewritten in the form:

\begin{eqnarray}
&\sigma _c^{DOS}&({0},\epsilon )=-\lim_{h\rightarrow 0}\frac{%
e^2sr\kappa }{8\eta }h\int_{-1/2}^{1/h+1/2}\frac{dn}{\sqrt{\epsilon +(2n+1)h}%
\sqrt{\epsilon +r+(2n+1)h}}=  \nonumber \\
&=&-\lim_{h\rightarrow 0}\frac{e^2sr\kappa }{8\eta }h{\bf \sum_{n=0}^{1/h}}%
\int_{-1/2}^{1/2}\frac{dx}{\sqrt{\epsilon +(2n+1/2+x)h}\sqrt{\epsilon
+r+(2n+1/2+x)h}}  \nonumber \\
&=&-\lim_{h\rightarrow 0}\frac{e^2sr\kappa }{8\eta }{\bf \sum_{n=0}^{1/h}}\ln 
{\frac{\sqrt{\epsilon +2nh+2h}+\sqrt{\epsilon +r+2nh+2h}}{\sqrt{\epsilon +2nh%
}+\sqrt{\epsilon +r+2nh}},}  \label{main1}
\end{eqnarray}
where $r=\frac{4\xi _c^2(0)}{s^2}$ is the dimensionless anisotropy parameter
of the Laurence-Doniach model (which is supposed to be small: $r\ll 1$), $s$
is the interlayer distance, $\eta $ is the gradient coefficient of the $2D$
Ginzburg-Landau theory. Here 
\begin{eqnarray}
\kappa  &=&{\frac{{-\psi ^{^{\prime }}({\frac 12}+{\frac 1{{4\pi \tau T}}})+{%
\frac 1{{2\pi \tau T}}}\psi ^{^{\prime \prime }}({\frac 12})}}{{\pi ^2[\psi (%
{\frac 12}+{\frac 1{{4\pi \tau T}}})-\psi ({\frac 12})-{\ \frac 1{{4\pi \tau
T}}}\psi ^{^{\prime }}({\frac 12})]}}}  \nonumber  \label{d20} \\
&\rightarrow &\cases{56\zeta(3)/\pi^4\approx0.691,&for $T\tau<<1$,\cr
8\pi^2(\tau T)^2/[7\zeta(3)]\approx9.384(\tau T)^2, &for $T\tau>>1$\cr}
\end{eqnarray}
is a function of $\tau T$ only \cite{BDKLV93}. It appears as the result of
summation of all DOS type diagrams.

Let us substitute the expression (\ref{main1}) in (\ref{delta}). Now for the
difference $\sigma _{c}^{DOS}({h},\epsilon )-\sigma _{c}^{DOS}(0,\epsilon )$
we may write the following formula, where the summation may be extended
until $N\longrightarrow \infty $ \ due to good convergence of the sum (the
n-th term of the sum is proportional to $n^{-3/2}$ for large $n$): 
\begin{eqnarray}
\Delta \sigma _{c}^{DOS}({h},\epsilon ) &=&\nonumber\\
&=&\frac{e^{2}sr\kappa }{8\eta}h{\bf \sum_{n=0}^{\infty }}\{\frac{1}{h} \ln {%
\frac{\sqrt{\epsilon +2nh+2h}+\sqrt{\epsilon +r+2nh+2h}}{\sqrt{\epsilon +2nh}%
+\sqrt{\epsilon +r+2nh}}}-  \nonumber \\
&&-\frac{1}{\sqrt{\epsilon +2nh+h}\sqrt{\epsilon +r+2nh+h}}\}  \label{main}
\end{eqnarray}

This expression is very suitable for numerical calculation to analyze
experimental data. It permits to obtain easily the asymptotic behavior of
magnetoconductivity in the case of non-weak fields (note the inaccuracy of
the analysis of this asymptotics in \cite{BDKLV93}). The case of very strong
fields $h\gg \max \{\epsilon ,r\}$, in contrast to \cite{BDKLV93}, becomes
now trivial for consideration: with logarithmic accuracy it is determined
just by the contribution of the first term in (\ref{main}) :

\begin{eqnarray}
\Delta \sigma _c^{DOS}(h\gg \max \{\epsilon ,r\})=\frac{e^2sr\kappa }{8\eta }%
\ln {\frac{\sqrt{2h}}{\sqrt{\epsilon }+\sqrt{\epsilon +r}}}.  \label{4*}
\end{eqnarray}
The further analysis of (\ref{main}) shows that for intermediate fields in
the temperature range of three dimensional fluctuations ($\epsilon \ll h\ll
r $) some nontrivial, typically 3D, behavior $\sim \sqrt{h}$ can be
obtained: 
\begin{eqnarray}
\Delta \sigma _c^{DOS}(\epsilon &\ll &h\ll r)=\frac{e^2sr\kappa }{16\eta }%
\sum_{n=0}^\infty \{\ln {\frac{1+\sqrt{\frac hr}\sqrt{2n+2}}{1+\sqrt{\frac hr%
}\sqrt{2n}}}-  \label{inter} \\
-\sqrt{\frac hr}\frac 1{\sqrt{2n+1}}\} &=&0.42\frac{e^2sr\kappa }{8\eta }%
\sqrt{\frac hr}.  \nonumber
\end{eqnarray}
The limit of weak fields $h\ll \epsilon $ turns out to be a too cumbersome
in the presentation (\ref{main}). Then the simplest way to obtain the
asymptotic is to use the Euler-McLaurin formula to transform the sum into
an: 
\begin{eqnarray}
\Delta \sigma _c^{DOS}({h\ll r},\epsilon )=\frac{e^2sr\kappa }{192\eta _2}%
\frac{2\epsilon +r}{\left[ \epsilon (\epsilon +r)\right] ^{3/2}}h^2.
\label{6**}
\end{eqnarray}

In addition to the DOS contribution it is necessary to take into account the
regular Maki-Thompson contribution, which in the case of weak fields takes
form \cite{BDKLV93}:

\begin{equation}
\Delta \sigma _{c}^{MT(reg)}(h\ll r,\epsilon )=\frac{e^{2}sr\tilde{\kappa}}{
192\eta}\frac{r}{\left[ \epsilon (\epsilon +r)\right] ^{3/2}}h^{2},
\end{equation}
where 
\begin{eqnarray}
\tilde{\kappa} &=&{\frac{{-\psi ^{^{\prime }}({\frac{1}{2}}+{\frac{1}{{4\pi
T\tau }}})+\psi ^{^{\prime }}({\frac{1}{2}})+{\frac{1}{{4\pi T\tau }}}\psi
^{^{\prime \prime }}({\frac{1}{2}})}}{{\pi ^{2}[\psi ({\frac{1}{2}}+{\frac{1 
}{{4\pi \tau T}}})-\psi ({\frac{1}{2}})-{\frac{1}{{4\pi \tau T}}}\psi
^{^{\prime }}({\frac{1}{2}})]}}}  \nonumber  \label{d25} \\
&\rightarrow &\cases{28\zeta(3)/\pi^4\approx0.3455 &for $T\tau<<1$,\cr
\pi^2/[14\zeta(3)]\approx0.5865 &for $T\tau>>1$\cr}
\end{eqnarray}
is another function of the impurities concentration only. We note that the
regular MT term has the same sign as the overall DOS contribution. In weak
fields it becomes noticeable only in 3D case $\epsilon \ll r$ and in the
dirty limit when $\tilde{\kappa}$ is comparable with $\kappa $.

The evaluation of the regular Maki-Thompson contribution to
magnetoconductivity may be done by the same procedure as in (\ref{main1})
for the analysis of non-weak fields :

\begin{eqnarray}
\Delta \sigma _c^{MT(reg)}(h,\epsilon ) &=&-\frac{e^2s\tilde{\kappa}}{4\eta
_2}h\sum_{n=0}^\infty \{\frac{\epsilon +(2n+1)h+r/2}{\sqrt{\epsilon +(2n+1)h}%
\sqrt{\epsilon +r+(2n+1)h}}-  \nonumber \\
&&-\frac 1h[\sqrt{\epsilon +(2n+3/2)h}\sqrt{\epsilon +r+(2n+3/2)h}-
\label{mtr} \\
&&\sqrt{\epsilon +(2n+1/2)h}\sqrt{\epsilon +r+(2n+1/2)h}]\}  \nonumber
\end{eqnarray}
For the 3D case in the region of intermediate fields it leads to 
\begin{eqnarray}
\Delta \sigma _c^{MT(reg)}(\epsilon \ll h\ll r) &=&\frac{e^2s\tilde{\kappa}}{%
4\eta }\sqrt{\frac hr}\sum_{n=0}\{\sqrt{2n+3/2}- \\
&&-\sqrt{2n+1/2}-\frac 1{2\sqrt{2n+1}}\}=0.02\frac{e^2s\tilde{\kappa}}{4\eta 
}\sqrt{\frac hr}.  \nonumber
\end{eqnarray}
One can see that the contribution $\Delta \sigma _c^{MT(reg)}(\epsilon \ll
h\ll r)$ can be of the same order as $\Delta \sigma _c^{DOS}(\epsilon \ll
h\ll r)$ (see (\ref{inter})) depending on the relation between parameters $%
0.1\tilde{\kappa}$ and $r\kappa $ ;  in dirty case it certainly has to be
taken into consideration.

The analysis of (\ref{mtr}) in the strong filed case ( $h\gg \max \{\epsilon
,r\}$ ) shows that

\begin{eqnarray}
\Delta &\sigma _c^{MT(reg)}&(h\gg \max \{\epsilon ,r\})= \nonumber \\
&=&\frac{\pi ^2e^2s\tilde{\kappa}}{128\eta }\frac 1h\{\epsilon (\epsilon
+r)+\frac 2{\pi ^2}(|\psi (1/4)|-|\psi (3/4)|-\frac{\pi ^2}4)r^2\} \nonumber\\
&=&\frac{\pi ^2e^2s\tilde{\kappa}}{128\eta } \cdot \frac{\epsilon (\epsilon
+r)+0.136r^2}h
\end{eqnarray}
and it evidently can be omitted in comparison with (\ref{4*}).

Let us analyze the results obtained starting from the 2D case $r\ll \epsilon 
$. The positive DOS contribution in magnetoconductivity growth as $H^2$ up
to $H_{c2}(\epsilon )$ (see (\ref{6**})) and then the crossover to a slow
logarithmic asymptotic takes place. One can notice that at $H\sim H_{c2}(0)$
the value of $\Delta \sigma _c^{DOS}(h\sim 1,\epsilon )=-\sigma
_c^{DOS}(0,\epsilon )$ what means the total suppression of the fluctuation
correction in such a strong field. The regular part of the Maki-Thompson
contribution does not manifest itself in this case.

In 3D case ($\epsilon \ll r$) the behaviour of magnetoconductivity is mainly
the same besides the existence of the intermediate region $\epsilon \ll h\ll
r$ with $\sqrt{h}$ dependence on magnetic field. Two crossovers take place
here: from $h^2$ to $\sqrt{h}$ (at $h\sim \epsilon $) and from $\sqrt{h}$ to 
$\ln {h}$ (at $h\sim r$) behaviour. It is important that in this case the
regular part of the Maki- Thompson contribution becomes to be comparable
with DOS contribution or even dominating on the latter (in the case of dirty
and very anisotropic superconductor). This domination anyway terminates at
strong fields $h\succeq r$ where $\Delta \sigma _c^{MT(reg)}({h},\epsilon )$
rapidly decreases $\sim \frac rh$ with the field increase  in contrast to
the robust $\Delta \sigma _c^{DOS}({h},\epsilon )\sim \ln \frac{{h}}r$ which
survives up to $h\sim 1$.

It worth to mention that the consideration presented above is valid not only
for c-axis magnetoconductivity but with minimal variations in the DOS
contribution coefficient, for any physical value like in-plane
magnetoconductivity, tunneling conductivity, NMR rate, Hall effect etc. In
these problems the DOS contribution certainly has to be treated side by side
with the AL and anomalous MT ones (if there are no special reasons for their
suppression, like in the case of c-axis transport for both of them, in the
case of NMR for AL contribution only, or in the case of d-pairing for the MT
one \cite{VBML98}).

It is important to stress that the suppression of DOS by magnetic field
contribution takes place very slowly. Such robustness with respect to the
magnetic field is of the same physical origin as the slow logarithmic
dependence of the DOS-type corrections on temperature. This differs strongly
the DOS contribution from Aslamazov-Larkin and Maki-Thompson ones \cite{AM78}%
, making the former noticeable in the wide range of temperatures (up to $%
\sim 2\div 3T_c$) and magnetic fields ($\sim H_{c2}(0)$). The scale of the
suppression of DOS contribution can be treated as the value of the pseudogap
observed in the experiments mentioned above \cite{VBML98}. It has the order
of $\Delta _{pseudo}\sim 2\div 3T_c$ for magnetoconductivity and NMR, $%
\Delta _{pseudo}\sim \pi T_c$ for tunneling and $\Delta _{pseudo}\sim \tau
^{-1}$ in optical conductivity.

{\bf Acknowledgments.} The authors are grateful to Prof. G.Balestrino, Prof.
W.Lang and Prof. O.Rapp for valuable discussions. This work was supported by
the NATO Collaborative Research Grant \#CRG 941187 and INTAS Grant \#
96-0452.


\begin{thebibliography}{99}
\bibitem{VBML98}  A.Varlamov, G.Balestrino, E.Milani, D.Livanov, {\em %
Advances in Physics}, in press (1998).

\bibitem{BMV95}  G.Balestrino, A.Varlamov, E.Milani, {\em JETP Letters} {\bf %
61}, 833 (1995).

\bibitem{BDKLV93}  V.Dorin, R.Klemm, A.A.Varlamov, A.Buzdin, D.Livanov, {\em %
Phys.Rev.}, {\bf B 48},12951 (1993).

\bibitem{Rapp96}  J.Axnas, W.Holm, Yu.Eltsev and O.Rapp, {\em Phys. Rev.
Lett.} {\bf 77}, 2280 (1996).

\bibitem{Heine98}  G.Heine, W.Lang et al., Phys.Rev.{\em Phys.Rev.}, in
press (1998).

\bibitem{ILVY93}  L. B. Ioffe, A. I. Larkin, A. A. Varlamov, L. Yu, {\em %
Phys. Rev.} {\bf B 47}, 8936 (1993).

\bibitem{ahl89}  A.G.Aronov, S.Hikami, A.I.Larkin, {\em Phys. Rev.Lett.} 
{\bf 62}, 965 (1989).

\bibitem{BD96}  A.Buzdin, V.Dorin {\em Fluctuation Phenomena in High
Temperature Superconductors}, edited by M.Ausloos and A.Varlamov, Kluewer
Academic Publishers, Dordrecht (1997).

\bibitem{AM78}  S.Ami, K.Maki {\em Phys. Rev.} {\bf B 18}, 4714 (1978).

\bibitem{CLRV96}  P.Carretta, D.Livanov, A.Rigamonti, A.Varlamov, {\em %
Phys.Rev.}, {\bf B 54}, R9682 (1993).
\end{thebibliography}
\end{document}